\documentclass[twocolumn,prb, superscriptaddress,showpacs]{revtex4-1}
\usepackage[T1]{fontenc}
\usepackage[latin9]{inputenc}
\usepackage{color}
\usepackage{bm}
\usepackage{amsmath}
\usepackage{amssymb}
\usepackage{graphicx}

\makeatletter
\@ifundefined{textcolor}{}
{%
 \definecolor{BLACK}{gray}{0}
 \definecolor{WHITE}{gray}{1}
 \definecolor{RED}{rgb}{1,0,0}
 \definecolor{GREEN}{rgb}{0,1,0}
 \definecolor{BLUE}{rgb}{0,0,1}
 \definecolor{CYAN}{cmyk}{1,0,0,0}
 \definecolor{MAGENTA}{cmyk}{0,1,0,0}
 \definecolor{YELLOW}{cmyk}{0,0,1,0}
}

\usepackage{hyperref}

\makeatother

\newcommand{\be}{\begin{equation}}
\newcommand{\ee}{\end{equation}}
\newcommand{\bea}{\begin{eqnarray}}
\newcommand{\eea}{\end{eqnarray}}
\newcommand{\ba}{\begin{eqnarray*}}
\newcommand{\ea}{\end{eqnarray*}}
\newcommand{\dagga}{{\phantom{\dagger}}}
\newcommand{\bR}{\mathbf{R}}

\newcommand{\bq}{\mathbf{q}}

\newcommand{\bk}{\mathbf{k}}

\newcommand{\bkp}{\mathbf{k'}}

\newcommand{\bRp}{\mathbf{R'}}

\newcommand{\dis}{\displaystyle}

\newcommand{\up}{\uparrow}
\newcommand{\down}{\downarrow}
\newcommand{\fract}[2]{\frac{\dis #1}{\dis #2}}

\newcommand{\eqn}[1]{(\ref{#1})}

\begin{document}

\title{Robust $s\pm$  Superconductivity in a Two-Band Hubbard-Fr{\"o}hlich Model of Alkali Doped Organics}

\author{Tao Qin}

\affiliation{International School for Advanced Studies (SISSA), and CNR-IOM Democritos
National Simulation Center, Via Bonomea 265, I-34136 Trieste, Italy }

\author{Michele Fabrizio}

\affiliation{International School for Advanced Studies (SISSA), and CNR-IOM Democritos
National Simulation Center, Via Bonomea 265, I-34136 Trieste, Italy }

\author{S. Shahab Naghavi}

\affiliation{International School for Advanced Studies (SISSA), and CNR-IOM Democritos
National Simulation Center, Via Bonomea 265, I-34136 Trieste, Italy }

\author{Erio Tosatti}

\affiliation{International School for Advanced Studies (SISSA), and CNR-IOM Democritos
National Simulation Center, Via Bonomea 265, I-34136 Trieste, Italy }

\affiliation{International Centre for Theoretical Physics (ICTP), Strada Costiera
11, I-34151 Trieste, Italy }

\email{E-mail: tosatti@sissa.it }

\begin{abstract}

The damaging effect of strong electron-electron repulsion on regular, electron-phonon 
superconductivity is a standard tenet. In spite of that, an increasing number of compounds such 
as fullerides and more recently alkali-doped aromatics exhibit 
superconductivity despite very narrow bands and very strong electron repulsion. Here, we explore 
superconducting solutions of a model Hamiltonian inspired by the electronic  structure of alkali 
doped aromatics. The model is a two-site, two-narrow-band metal with a single intersite phonon, leading 
to attraction-mediated, two-order parameter superconductivity.  On top of that, the model includes 
a repulsive on-site Hubbard $U$, whose effect on the superconductivity we study. Starting within mean 
field, we find that $s \pm$ superconductivity is the best solution surviving the presence of $U$, whose effect is 
canceled out by the opposite signs of the two order parameters.  The correlated Gutzwiller study that 
follows is necessary because without electron correlations the superconducting state would in this
model be superseded by an antiferromagnetic insulating state with lower energy. The Gutzwiller correlations  
lower the energy of the metallic state, with the consequence that the $s \pm$ superconducting state 
is stabilized and even strengthened for small Hubbard $U$.
 
\end{abstract}

\pacs{74.20.-z, 74.10.+v, 74.20.Mn, 74.70.Kn}

\maketitle

\section{Introduction}
The long time search for superconductivity in electron doped organic
molecular crystals has recently included common polycyclic aromatic
hydrocarbons (PAHs) such as picene, coronene, phenanthrene and others,
where evidence for doping-induced diamagnetic fractions has
been reported, suggesting superconductivity with properties yet to
be established.~\cite{PAH-SC1,PAH-SC2,PAH-SC3,PAH-SC4, wang2011} This represents an interesting research
direction, both because of the desirability of cheap, light and environment
friendly new superconductors, and of the potential novelties implied
by the added molecular complexity. One is faced however with riddles, 
including very basic ones such as the structure and stoichiometry of the 
unknown superconducting compound fractions. What is the compound 
crystal structure, what is the variety of phases which may occur, and what 
is the reason why superconductivity is mostly reported for three nominally added alkalis
are wide open questions. Moreover the interplay of strong correlations and electron-phonon,
both expected to be strong, is unclear.

While we must await further experiments and reliable data to address many of these questions,
theoretical modeling can help clarifying at least some
of them. 
The electron bandwidths
$W$ of hypothetical $\left(\mathrm{X}\right)^{3+}\left(\mathrm{PAH}\right)^{3-}$
compounds ($\mathrm{X}$= three alkalis, or one trivalent metal such
as La) have been recently calculated, ~\cite{Boeri, Kosugi09, aoki, spaniards, giovannetti, calandra, naghavi, yan2013} 
and found to be comparable to, generally narrower than, the estimated value of the intra-molecular
Coulomb repulsion $U$.~\cite{nomura} While that suggests strong
electron correlations, 
with possible
proximity of Mott insulating states and related
phenomena~\cite{giovannetti} akin to those invoked for systems such
as cuprates, $\kappa\mathrm{(ET)_2}X$ organics, and fullerides,\cite{cuprate-review,kanoda,capone09}
no clear evidence in this direction,
such as e.g., a large magnetic susceptibility,
has actually emerged so far.

On the other hand, a very substantial intra-molecular and, remarkably,
\textit{inter}-molecular electron phonon coupling strength has been calculated.~\cite{calandra}
Thus, if correlations could  be canceled, some kind of BCS-type superconducting state 
might be realized. Lacking reliable experimental information, a 
variety of possible  crystal structures of alkali doped aromatics 
are currently being addressed by density functional theory (DFT) total energy studies
including our own~\cite{naghavi, yan2014, naghavi14} where, depending on 
the unit-cell structure, both insulating and metastable metallic phases emerge. 
In a hypothetical metallic phase of La-phenanthrene~\cite{naghavi},
which we 
adopt here as
our prototype, a simplified model Hamiltonian was extracted.
It is a two-site, two-narrow-band model, with a large Fr{\"o}hlich  electron-phonon coupling to a single {\it inter}-site phonon, 
and an on-site electron-electron repulsive Hubbard $U$.  With these ingredients, the model is referred to as 
a Hubbard-Fr{\"o}hlich two-band model.

We regard this kind of model of rather general interest because of a multiplicity of reasons. Two molecules per cell,
generally stacked in a herringbone fashion, is a widespread structural motif in doped polycyclic aromatic hydrocarbon synthetic
metals. That kind of structure leads to two narrow and often partly degenerate  LUMO+1 bands, which become half-filled at
the trivalent electron doping of wider interest~\cite{PAH-SC2}. The partial degeneracy is, as we observed earlier,
effectively lifted by a dimerizing distortion, which brings together pairs of molecules. A zone-boundary intermolecular phonon
enacting that displacement  thus exhibits the strongest electron-phonon coupling near the Fermi level, and is adopted as
the main ingredient of the model~\cite{naghavi}.  Finally, because the bands are narrow and the Coulomb electron-electron
repulsion cannot be considered negligible~\cite{spaniards,naghavi,Boeri,calandra,casula,aoki,giovannetti}.  Ignoring intermolecular interactions, the repulsive Coulomb effects are represented
by an intra-site Hubbard $U$.   

In this paper, we explore and discuss the superconducting solution of this Hubbard-Fr{\"o}hlich two-band model, 
where superconductivity may arise driven by phonon attraction, but has to reckon with the repulsive Hubbard $U$. 
Due to the intersite symmetry, we find that 
two BCS gaps with opposite sign effectively cancel the effect of Coulomb repulsion in an $s \pm$-wave
electron-phonon superconducting state. Within 
uncorrelated mean field theory therefore,  $s\pm$ phonon
superconductivity survives unscathed up to large Hubbard $U$ values, 
where regular $s$-wave superconductivity in a 
single-site 
Hubbard-Holstein model would be hopelessly
suppressed. 
To further 
test the robustness of this two-gap 
superconducting state against the alternative possibility of an insulating magnetic solution, always present and in fact prevailing over superconductivity if correlations are ignored, 
we include Gutzwiller correlations in our model solution. Upon inclusion of correlations, superconductivity
survives as the most stable phase up to a threshold value $U \le U_c$ of electron-electron 
repulsion.

\section{The model}\label{THE MODEL}

\begin{figure}
\includegraphics[scale=0.25,width=1.\linewidth]{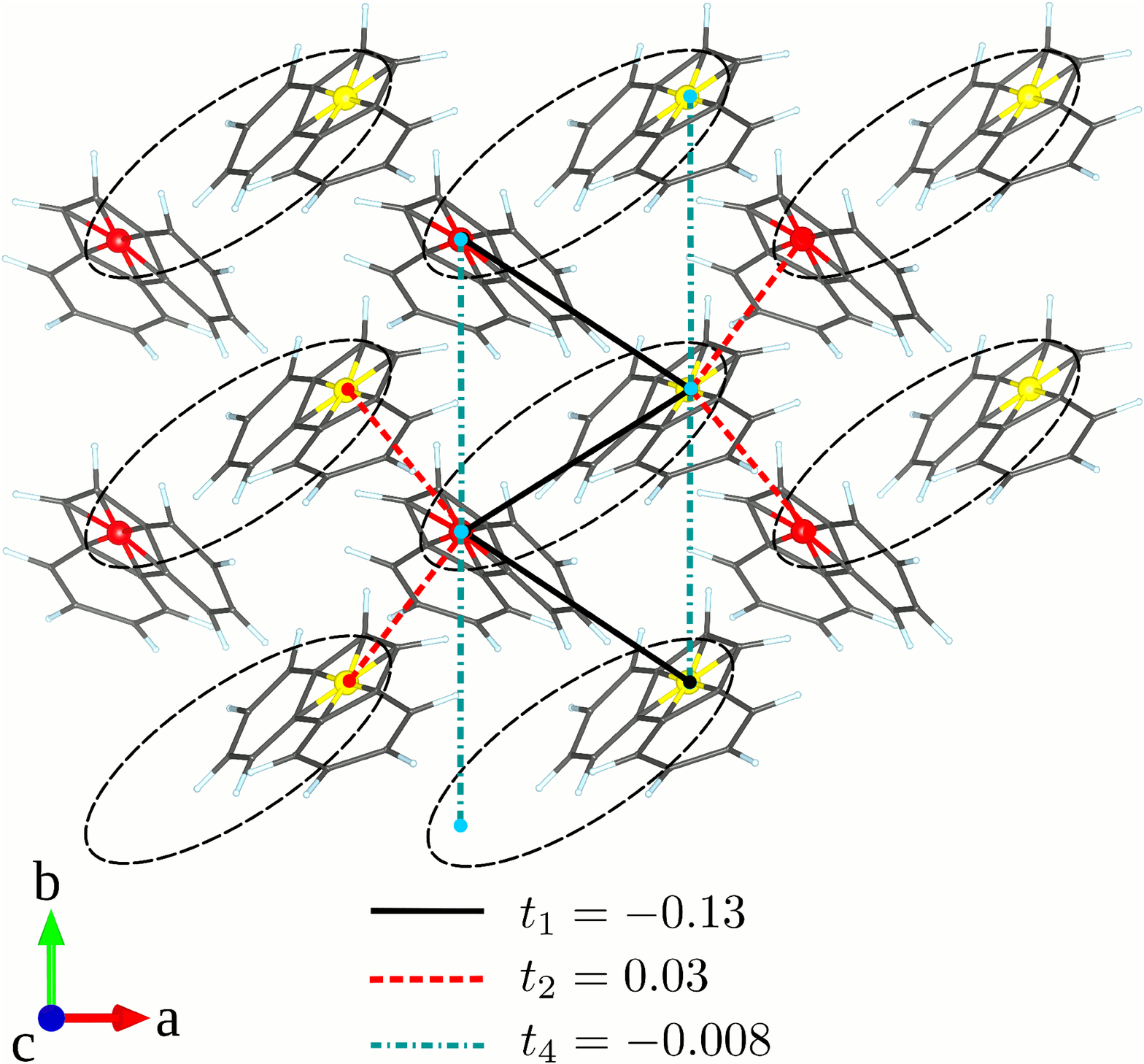}

\caption{\label{fig:phonon}(Color online) Schematic molecular lattice model, adapted from Ref.~\onlinecite{naghavi}. Inter-molecular 
electron hopping matrix elements $t_1$, $t_2$, $t_4$ are marked. The inter-site dimerizing phonon between sites connected
by the black line is responsible for the superconductivity in the Hubbard-Fr{\"o}hlich model in Eq.\,\eqref{Hamiltonian}. }
\end{figure}

\begin{figure}
\includegraphics[scale=0.4,width=1.\linewidth]{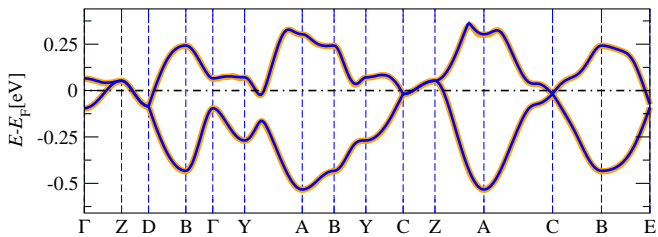}

\caption{\label{fig:HF}(Color online) The two half-filled bands used for the modeling, qualitatively representing LUMO+1 derived
bands in a generic three-electron-doped polycyclic aromatic hydrocarbon. The specific form and parameters are from DFT results 
(black line) and their Wannier parametrization (orange line) obtained for hypothetical La-phenthrene in Ref.~\onlinecite{naghavi}. }
\end{figure}

We start off with the two-band tight-binding model recently proposed
in Ref.~\onlinecite{naghavi}. The assumed three-dimensional lattice 
sketched in Fig. \,\ref{fig:phonon} has $\mathrm{P2_{1}}$ symmetry, typical of many even if not all pristine
PAHs~\cite{naghavi, casula, aoki2}, with two equivalent sites per
cell. An important symmetry element is the screw axis, which transforms
one molecule onto the other, through a rotation accompanied by a fractional
translation. Each site is endowed with 
a single nondegenerate 
orbital, representing
the second lowest unoccupied molecular orbital (LUMO+1) of the neutral molecule. With an average
of three electrons per molecule donated by electropositive atoms,
(not included in the model), each molecular LUMO is completely filled and can be ignored, 
so that the LUMO+1 derived states that are precisely the half filled band that must be treated.
Electrons 
in this orbital
hop 
between sites with matrix elements 
indicated in Fig. 1, modeled for specificity after calculations performed for a representative
hypothetical metallic phase of La-phenanthrene, giving rise to a half-filled LUMO+1
derived pair of bands~\cite{naghavi} shown in Fig.2. As seen in  this figure 
the screw axis symmetry causes an important 
partial degeneracy on the Brillouin zone boundary  -- the two bands \textquotedbl{}sticking\textquotedbl{}
together~\cite{heine} -- near the Fermi level. Two additional ingredients of the model are an intra-site Coulomb 
\textquotedbl{}Hubbard\textquotedbl{} $U$, and a \textquotedbl{}Fr{\"o}hlich\textquotedbl{} coupling of
electronic states to an inter-site phonon, whose key feature is a \textquotedbl{}dimerizing\textquotedbl{}
character. 
A dimerizing displacement brings nearest molecules closer to form pairs, and
is precisely such as to remove the screw axis, thus splitting
the band degeneracy near Fermi level.~\cite{naghavi}  We note here by analogy that the ability to split
a band degeneracy near the Fermi level (that of the $\sigma$ bonding band
top) is the basic reason why the famous $E_{g}$ phonon is so very effectively
driving superconductivity in $\mathrm{MgB_{2}}$.~\cite{MgB2} 
The Hubbard-Fr{\"o}hlich model Hamiltonian is 
\begin{equation}
\mathcal{H}=\mathcal{H}_{0}+\mathcal{H}_{\mathrm{ph}}+\mathcal{H}_{\mathrm{el-ph}}+\mathcal{H}_{\mathrm{U}},\label{Hamiltonian}
\end{equation}
where electron hopping is 
\begin{equation}
\mathcal{H}_{0}=\sum_{\bk \sigma}\left(\begin{array}{cc}
c_{1\bk \sigma}^{\dagger} & c_{2\bk\sigma}^\dagger\end{array}\right)\left(\begin{array}{cc}
t_{\bk} & t_{\bk}^{12}\\
t_{\bk}^{12\ast} & t_{\bk}
\end{array}\right)\left(\begin{array}{c}
c_{1\bk\sigma}^\dagga\\
c_{2\bk\sigma}^\dagga
\end{array}\right),\label{electron-hopping}
\end{equation}
where $c^\dagger_{1(2)\bk\sigma}$ creates a spin-$\sigma$ electron at 
momentum $\bk$ in molecule 1(2). $\mathcal{H}_0$ in Eq. 
\eqn{electron-hopping} gives rise to the half filled bands of Fig.\,\ref{fig:HF}. While we believe 
that many of the results to be derived later have a sufficient level of generality, the matrix elements in 
\eqn{electron-hopping} are borrowed for specificity from an electronic structure calculation in Ref.~\onlinecite{naghavi} and are 
\bea
t_\bk &=& 2\Big(t_{4}\,\cos k_{y}+t_{5}\,\cos k_{z}+t_{6}\,\cos\left(k_{x}+k_{z}\right)\Big),\label{diagonal}\\
t^{12}_\bk &=& \Big(1+\text{e}^{-ik_{y}}\Big)\Big(t_{1}+t_{2}
\,\text{e}^{-ik_{x}}+t_{3}\,\text{e}^{ik_{z}}\Big)\nonumber\\
&& \equiv - \text{e}^{-i\theta_\bk}\;\tau_\bk,\label{off-diagonal}
\eea
where $t_{1}\simeq-0.13$ eV, $t_{2}\simeq 0.03$ eV,
$t_{3}\simeq 0.07$ eV, $t_{4}\simeq -0.008$ eV, $t_{5}\simeq 0.014$ eV, and finally $t_{6}\simeq 0.013$ eV.\cite{naghavi}
Recent calculations for K$_3$-phenanthrene,~\cite{naghavi14} a system where superconductivity has been observed~\cite{wang2011} 
lead to a LUMO+1 band structure that is 
similar, although different in the details.

As in Ref.~\onlinecite{naghavi}, 
and as indicated in Fig. 1,  
we only include in the model the inter-site phonon 
that modulates the hopping between molecule 1 and 2 along the 
$b$ direction, where the hybridization is stronger. This phonon 
has a dispersion 
\begin{equation}
\mathcal{H}_{\mathrm{ph}}=\sum_{\bq}\,\fract{\omega_{\bq}}{2}\;\Big(p_{\bq}\,p_{-\bq}
+x_{\bq}\,x_{-\bq}\Big),
\label{inter-site phonon}
\end{equation}
and is Fr{\"o}hlich coupled to the conduction electrons via  
\begin{align}
\mathcal{H}_{\mathrm{el}-\mathrm{ph}}= & \sum_{\bk,\bq,\sigma}x_{-\bq}\,\left(\gamma_{\bk+\bq}+
e^{i\bq\cdot\mathbf{b}/2}\gamma_{\bk}\right)\,c_{1\bk\sigma}^{\dagger}c_{2\bk+\bq\sigma}^\dagga\nonumber \\
 & +x_{-\bq}\left(\gamma_{-\bk}+e^{i\bq\cdot\mathbf{b}/2}\gamma_{-\bk-\bq}\right)\, c_{2\bk\sigma}^{\dagger}c_{1\bk+\bq\sigma}^\dagga.\label{el-ph}
\end{align}
Finally, the Hubbard repulsion is 
\begin{equation}
\mathcal{H}_{\mathrm{U}}=U\,\sum_{\bm{R}}\, n_{1\bm{R}\uparrow}\,n_{1\bm{R}\downarrow}+n_{2\bm{R}\uparrow}\,n_{2\bm{R}\downarrow}.
\end{equation}

At half-filling, the Hamiltonian \eqn{Hamiltonian} has in principle two competing instabilities: 
(i) antiferromagnetic 
insulator, 
the two molecules in the unit cell with opposite spin-$\frac{1}{2}$ polarization; (ii) phonon-mediated superconductivity. 
Antiferromagnetism is frustrated by the $k$-dependent diagonal elements in the hopping matrix of Eq. \eqn{electron-hopping}, 
and can be expected to prevail only above a threshold value of the repulsion $U$. The phonon-mediated Cooper instability only 
requires a finite density of states at the chemical potential and, since the pairing channel is intermolecular, it might be able to escape 
the intramolecule repulsion $U$. This qualitative reasoning leads us to expect that superconductivity might occur below a critical $U_c$,
and antiferromagnetism above that. This simple-minded expectation will be explored and substantiated by calculations in the following sections.

\section{Mean field solution}

The simplest tool to search for instabilities in an interacting electron model is the Hartree-Fock approximation. In our case, this is complicated by the retardation of the phonon-mediated electron-electron interaction. As in BCS theory, we shall neglect retardation and approximate the phonon-mediated interaction $\mathcal{H}_\text{el-el}$ by an instantaneous attraction that we will assume to act between electrons closer to the Fermi energy than a cutoff energy of order the Debye frequency.  

The first mean-field step is diagonalizing the noninteracting Hamiltonian $\mathcal{H}_0$ in Eq. \eqn{electron-hopping}. This is done by applying the unitary  transformation [see Eqs. \eqn{diagonal} and \eqn{off-diagonal}]
\bea
c_{g\bk\sigma}^{\dagger}&=&\fract{1}{\sqrt{2}}\;\Big(
c_{1\bk\sigma}^{\dagger}+\text{e}^{i\theta_{\bk}}\;c_{2\bk\sigma}^{\dagger}\Big),\label{c-1}\\
c_{u\bk\sigma}^{\dagger} &=& \fract{1}{\sqrt{2}}\;\Big(c_{1\bk\sigma}^{\dagger}-\text{e}^{i\theta_{\bk}}\;c_{2\bk\sigma}^{\dagger}\Big),\label{c-2}
\eea
which leads to  
\be
\mathcal{H}_{0}=\sum_{\bk\sigma}\, \xi_{g\bk}\,c_{g\bk\sigma}^{\dagger}\,c_{g\bk\sigma}^\dagga+\xi_{u\bk}\,c_{u\bk\sigma}^{\dagger}\,c_{u\bk\sigma}^\dagga,\label{H_0-diagonal}
\ee
where $\xi_{g\bk} = t_\bk - \tau_\bk -\mu$ and 
$\xi_{u\bk} = t_\bk + \tau_\bk -\mu$ are the energies measured with respect to the chemical potential $\mu$. 

Since superconductivity acts between time-reversed partners, pairing 
must be intra-band. We thus concentrate on the spin-singlet pair-creation operators
\bea
\Delta^\dagger_{g\bk} &=& c^\dagger_{g\bk \up}c^\dagger_{g -\bk\down}
+ c^\dagger_{g -\bk \up}c^\dagger_{g \bk\down},\label{Delta-g}\\
\Delta^\dagger_{u\bk} &=& c^\dagger_{u\bk \up}c^\dagger_{u -\bk\down}
+ c^\dagger_{u -\bk \up}c^\dagger_{u \bk\down},\label{Delta-u}
\eea
of each band, with energies $2\xi_{g\bk}$ and $2\xi_{u\bk}$, respectively. 
The Fr\"ohlich-type of electron-phonon coupling, Eq. \eqn{el-ph}, can generate 
either an inter-molecular pairing (see Sec. \eqref{appendix} for details.)

\begin{align}
\propto\, &\; -\big(c^\dagger_{1\bk\up}c^\dagger_{2-\bk\down}\,c^\dagga_{1-\bkp\down}c^\dagga_{2\bkp\up}  + 
c^\dagger_{2\bk\up}c^\dagger_{1-\bk\down}\,c^\dagga_{2-\bkp\down}c^\dagga_{1\bkp\up}\big)
\nonumber\\
& \sim  -\big(\Delta^\dagger_{g\bk}-\Delta^\dagger_{u\bk}\big)\big(\Delta^\dagga_{g\bkp}
-\Delta^\dagga_{u\bkp}\big),\label{inter-molecule-pairing}
\end{align}
or a pair hopping term
\begin{align}
\propto\, &\; -\big(c^\dagger_{1\bk\up}c^\dagger_{1-\bk\down}\,c^\dagga_{2-\bkp\down}c^\dagga_{2\bkp\up}  + 
c^\dagger_{2\bk\up}c^\dagger_{2-\bk\down}\,c^\dagga_{1-\bkp\down}c^\dagga_{1\bkp\up}\big) 
\nonumber\\
& \sim  -\big(\Delta^\dagger_{g\bk}+\Delta^\dagger_{u\bk}\big)\big(\Delta^\dagga_{g\bkp}
+\Delta^\dagga_{u\bkp}\big), \label{pair-hopping}
\end{align}
which, when combined, justify the following expression for the phonon-mediated attraction that we shall 
consider hereafter:
\begin{equation}
\mathcal{H}_{\mathrm{ep-eff}}=-\fract{g_{\ast}}{2V}\sum_{\bk\bkp}\,s_{\bk}\,
s_{\bkp}\Big(\Delta_{g\bk}^{\dagger}\,\Delta_{g\bkp}^\dagga +\Delta_{u\bk}^{\dagger}\,
\Delta_{u\bkp}^\dagga\Big).\label{el-el-phon-media}
\end{equation}  
Here $g_{\ast}$ is the effective attractive potential, which is of the order of the square of the typical electron-phonon coupling constant $\gamma$, 
see Eq. \eqn{el-ph}, divided by the typical phonon frequency. As mentioned, 
we neglect retardation but introduce a function $s_\bk$ which
is +1 if $\left|\xi_{g(u) \bk}\right|\le\hbar\omega_{D}$
with $\omega_{D}$ the typical phonon frequency, and zero otherwise.

 The Hubbard repulsion, once projected onto the intra-band singlet Cooper channels, reads as 
\begin{equation}
\mathcal{H}_{\mathrm{U}}=\frac{U}{8V}\sum_{\bk\bkp}\left(\Delta_{g\bk}^{\dagger}+
\Delta_{u\bk}^{\dagger}\right)\left(\Delta_{g\bkp}+\Delta_{u\bkp}\right).\label{H-U-projected}
\end{equation}

We solve the Hamiltonian $\mathcal{H}=\mathcal{H}_{0}+\mathcal{H}_{\mathrm{ep-eff}}+\mathcal{H}_{\mathrm{U}}$ 
within the Hartree-Fock approximation (HF). Assuming the two order parameters 
\bea
\bar{\Delta}_{g\bk}^{\ast} &=& \left\langle \Delta_{g\bk}^{\dagger}\right\rangle, \\
\bar{\Delta}_{u\bk}^{\ast}&=& \left\langle \Delta_{u\bk}^{\dagger}\right\rangle ,
\eea
the gap equations are 
\begin{equation}
\bar{\Delta}_{g\left(u\right)\bk}=\fract{D_{g\left(u\right)\bk}}{E_{g\left(u\right)\bk}}
\; \bigg[2f\big(E_{g\left(u\right)\bk}\big)-1\bigg],\label{eq: order}
\end{equation}
where $f(E)$ is the Fermi-Dirac distribution at temperature $T$, and 
\bea
D_{g\left(u\right)\bk}&=&\frac{1}{V}\sum_{\bkp}\Bigg[
\frac{U}{4}\,\Big(\bar{\Delta}_{g\left(u\right)\bkp}+
\bar{\Delta}_{u\left(g\right)\bkp}\Big)\nonumber\\
&& \qquad \qquad -g_{\ast}\, s_{\bk}\, s_{\bkp}
\; \bar{\Delta}_{g\left(u\right)\bkp}\Bigg],\label{D_g(u)}\\
E_{g\left(u\right)\bk}&=& \sqrt{\xi_{g\left(u\right)\bk}^{2}+D_{g\left(u\right)\bk}^{2}}\, ,
\label{E_g(u)}
\eea
with the additional assumption that $\bar{\Delta}_{g\left(u\right)\bk}$ 
is real. We also need to fix the chemical potential $\mu$ so that the density corresponds to one electron per site, 
which brings about another self-consistency equation besides \eqn{eq: order}. 

Before discussing the solution of the self-consistency equations, it is worth remarking that, if $U>4g_\ast$, as 
is expected to be the case, the whole interaction, Eq. \eqn{H-U-projected} plus Eq. \eqn{el-el-phon-media}, 
is repulsive everywhere in momentum space, although its value jumps from $U-4g_\ast$ when $s_\bk\,s_\bkp =1$ up to $U$ when 
 $s_\bk\,s_\bkp = 0$. In spite of that overall repulsion, superconductivity can still as we shall see be stabilized 
by two conspiring facts: (1) the high-energy screening of the Hubbard $U$ which results into an effectively  
lower repulsion, which in the long-range case results in the so-called Coulomb pseudo-potential felt by the 
electrons close to Fermi level;\cite{Anderson&Morel}
(2) the opposite sign which can be chosen by the two order parameters $\bar{\Delta}_{g\bk}$ and $\bar{\Delta}_{u\bk}$ 
radically reducing the strength of the Hubbard repulsion, see Eq. \eqn{H-U-projected}, an $s_\pm$-wave symmetry 
similar to the Suhl-Kondo scheme for band-overlapping superconductors.\cite{Suhl-PRL,Kondo-SC} 
In fact, we note that, among the two pairing channels Eqs. \eqn{inter-molecule-pairing} and 
\eqn{pair-hopping} that can be stabilized by the electron-phonon coupling Eq. \eqn{el-ph}, only the former, 
which corresponds to an inter-molecule spin-singlet pairing,  
is not hindered by the Hubbard repulsion Eq. \eqn{H-U-projected}, thus naturally explaining the reason 
of the sign difference between $\bar{\Delta}_{g\bk}$ and $\bar{\Delta}_{u\bk}$. In other words, 
it is crucial for the stabilization of superconductivity despite the Hubbard repulsion that electrons couple 
to phonons in a Fr\"ohlich's inter-site rather than Holstein's intra-site fashion, i.e. by a phonon-modulated 
hopping rather than by a phonon-induced  charge attraction.

\subsection{Hartree Fock Results}

\begin{figure}
\includegraphics[scale=0.5,width=1.\linewidth]{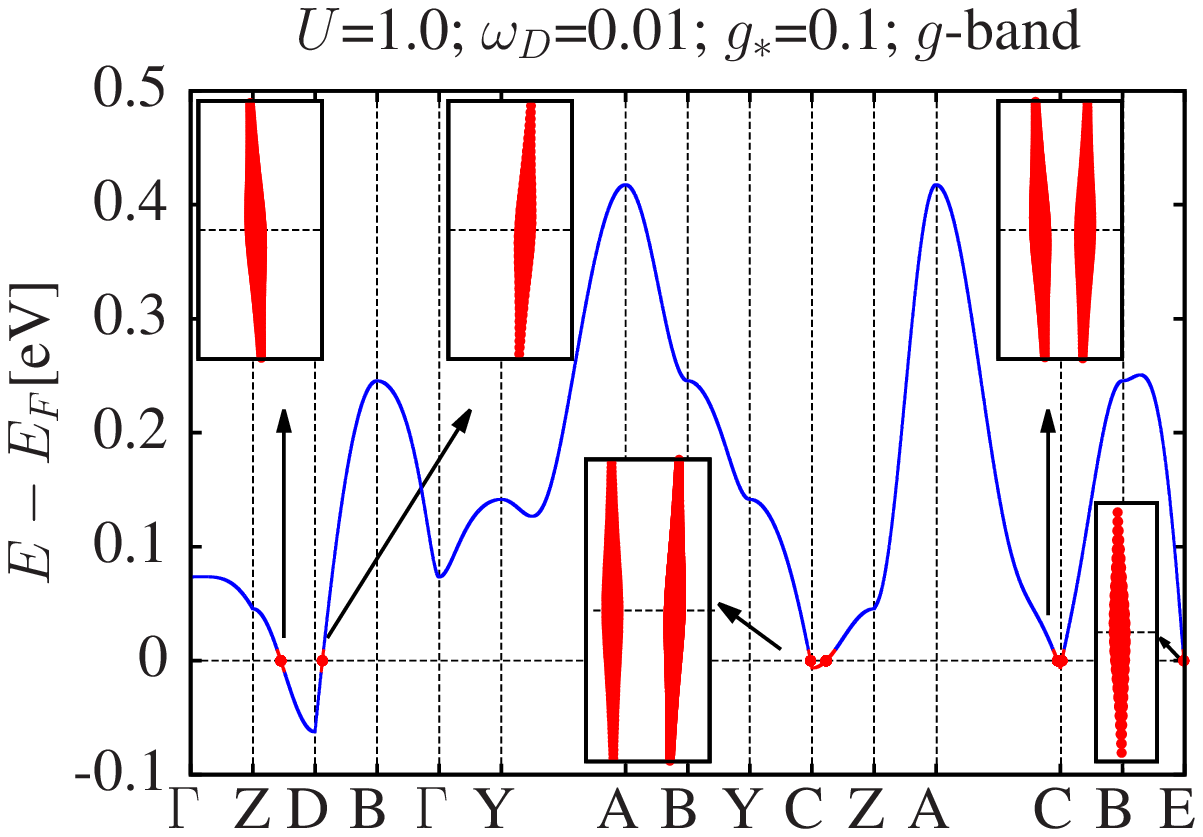}
\includegraphics[scale=0.5,width=1.\linewidth]{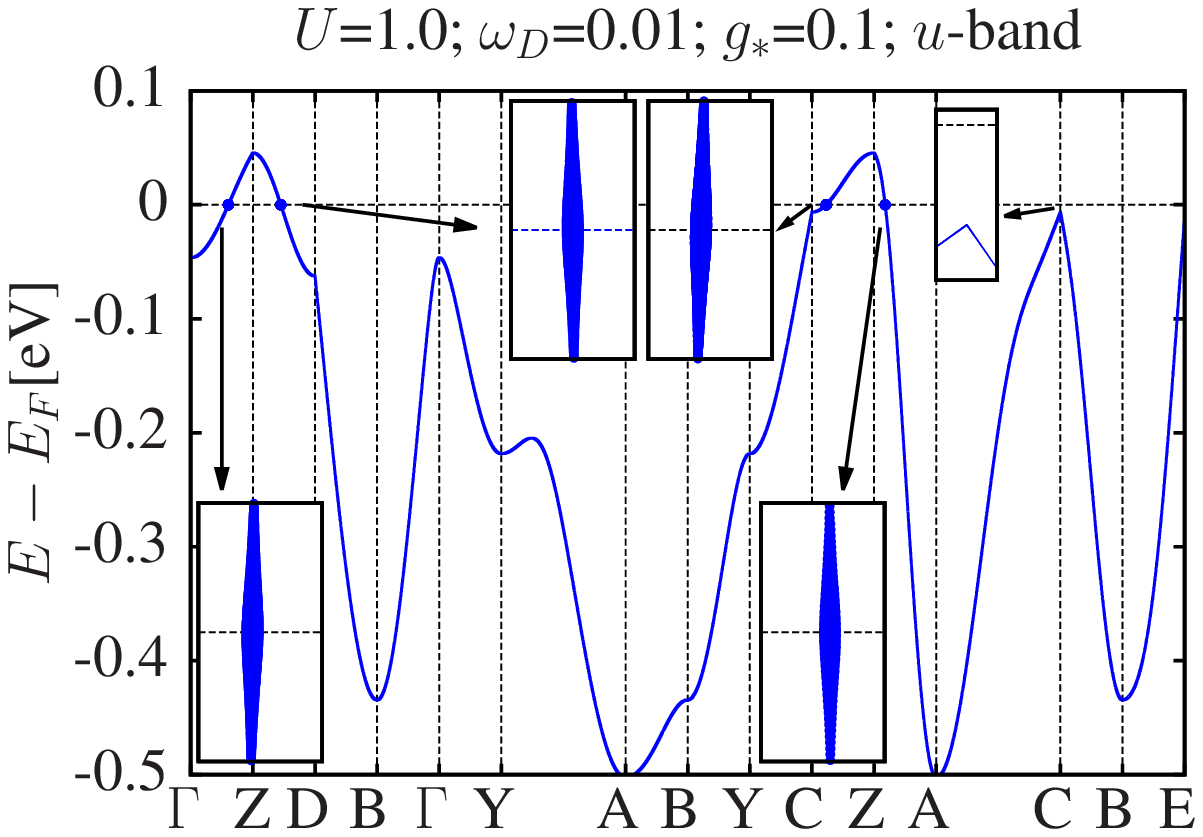}

\caption{\label{MFA_gaps}(Color online) Energy bands and gap parameters in the main symmetry directions in the $\bm{k}$ space at zero
temperature.  The insets show the magnified gap parameters near the Fermi level: width proportional to amplitude, blue and 
red colors indicates positive and negative sign.}
\end{figure}

\begin{figure}
\includegraphics[scale=0.5,width=1.\linewidth]{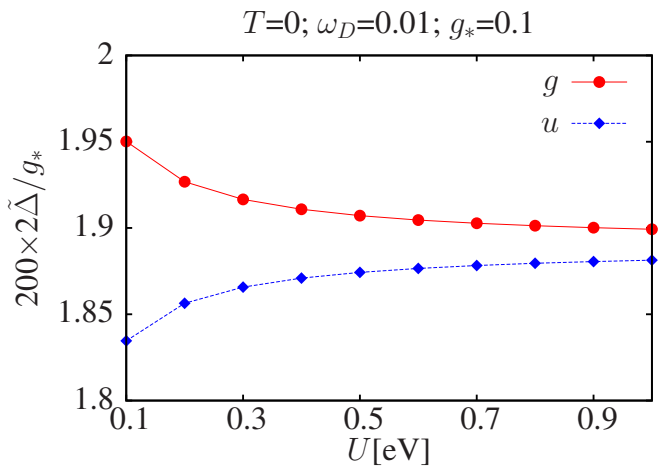}

\caption{\label{MFA_gapU}(Color online)  $T$=0 mean-field energy gap magnitudes are robust for increasing Hubbard
$U$.}
\end{figure}

\begin{figure}
\includegraphics[scale=0.5,width=1.\linewidth]{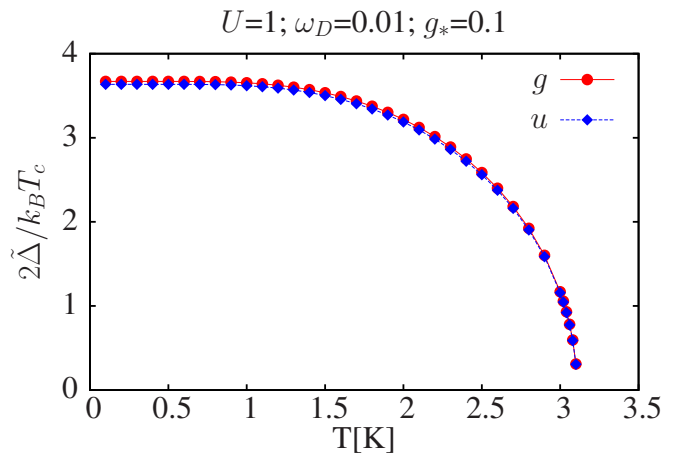}

\caption{\label{MFA_gapT}(Color online) Mean-field temperature dependence of energy gaps. Parameters such as those of 
Ref.~\onlinecite{naghavi}  would lead to an estimated $T_c \sim$ 3 K.}
\end{figure}

We solved numerically the Hartree-Fock self-consistency equations, Eq.\,\eqn{eq: order} with the condition that fixes 
the chemical potential, using the tight-binding parameters extracted in Ref.~\onlinecite{naghavi}, 
and reasonable estimates of the electron-phonon coupling $g_\ast = 100$ meV and of the cut-off Debye frequency 
$\hbar\omega_D = 10$ meV.\cite{Boeri,calandra,casula,Girlando}  As a parameter, we considered a variable electron repulsion 
$U$ below 1.0 eV,\cite{nomura} still reaching and even surpassing the bandwidth, so as to assess its importance 
in connection with superconductivity.  As anticipated, see Fig.\,\ref{MFA_gaps}, the two gaps acquire
opposite sign $\bar{\Delta}_{g\bm{k}}\bar{\Delta}_{u\bm{k}}<0$ in
the main symmetry directions, so that the cancellation between $\bar{\Delta}_{g\bm{k}}$
and $\bar{\Delta}_{u\bm{k}}$ preserves superconductivity in spite of a substantial repulsion. In addition, both gaps 
change sign at an energy equal to the cutoff $\hbar\omega_D$; the standard manifestation of the the high-energy 
screening. 
\cite{Anderson&Morel} As a result, 
the gap values (Fig.\,\ref{MFA_gapU}) are remarkably insensitive to the growth of $U$ in a broad range. In Figs. \ref{MFA_gapU} and 
\ref{MFA_gapT} the 
even and odd superconducting gaps $\tilde{\Delta}_{g}$ and $\tilde{\Delta}_{u}$ are defined as averages over the Fermi surfaces.

In Fig.\,\ref{MFA_gapT}, we show instead the 
gaps $\tilde{\Delta}_{g}$ and $\tilde{\Delta}_{u}$ as function of temperature. With the assumed parameters we 
can estimate  $T_c \sim  3$ K, in fact not dissimilar to the experimental ones.\cite{PAH-SC4}  
We should probably regard that
order of magnitude
agreement as coincidental, both the model and the approximations being rather generic.  

\section{Gutzwiller Correlations and Gutzwiller Approximation}
\label{Gutzwiller Correlations}

In the previous section, we found that a variational BCS state can be stabilized 
within the Hartree-Fock approximation in spite of a relatively strong Coulomb repulsion, thanks to an order parameter that develops opposite sign in the two bands, an $s_\pm$-wave symmetry\cite{Suhl-PRL,Kondo-SC}. This sign change suppresses the onsite amplitude of the pair wave function, thus reducing the energy cost of the Hubbard $U$. As we already mentioned [see Eq. \eqn{inter-molecule-pairing}], the property  
$\bar{\Delta}_{g\bm{k}}\,\bar{\Delta}_{u\bm{k}}<0$ is in reality a characteristic of an inter-site pairing 
\be
c^\dagger_{1\bk\up}c^\dagger_{2-\bk\down} + c^\dagger_{2-\bk\up}c^\dagger_{1\bk\down} 
\sim \Delta_{g\bk}^\dagger - \Delta_{u\bk}^\dagger,\label{1-2-pairing}
\ee
as opposed to an on-site one. 
In other words, the sign difference is brought here by the 
pairing mechanism itself rather than by the competition with onsite repulsion. 
As a matter of fact, the latter may actually strengthen pairing. In fact, as $U$ increases, the time each pair 
of neighboring molecules spend in the configuration where both are singly occupied increases, 
leaving enough time for the molecules to couple into an inter-site spin-singlet thus gaining 
electron-phonon energy before the electrons escape. 

The Hartree-Fock approximation is not fully able to grasp this repulsion-reinforced pairing, exhibiting 
a superconducting gap that does not grow but rather saturates for large $U$ (see Fig.\,\ref{MFA_gapU}). 
This 
limitation 
of  the Hartree-Fock approximation does not come as a surprise, since 
the method is not reliable when the interaction is comparable or even larger than the bandwidth. 
This uncertainty becomes crucial if 
we must compare the superconducting state energy with other possible ground states that on the contrary 
take advantage of a greater $U$, most notably an antiferromagnetic insulator with molecule 1 spin-polarized 
opposite to molecule 2. Physically, large $U$ tends to Mott localize the charge by suppressing configurations 
where two electrons occupy the same LUMO+1 molecular orbital. In order not to waste too much kinetic 
and ionic-potential energy, the electrons must coordinate among each other so to avoid sharing the same 
molecule during their motion. This electron self-organization occurs at large $U$ 
in correspondence with
charge localization.  Antiferromagnetism is but a strategy to synchronize electron motion, forcing nearby molecules 
to be occupied by opposite spin electrons which can therefore exchange.  Antiferromagnetism has indeed 
been shown to arise as the lowest-energy solution within density functional theory calculations of alkali-doped aromatics~\cite{giovannetti}. 
Our point here is that the true system has other strategies at its disposal. 
In fact, efficient correlations avoiding double occupancy can also develop within an overall singlet 
and metallic ground 
state, including the above $s_\pm$ superconducting state stabilized by the Fr\"ohlich's electron-phonon coupling.   
In order to explore that possibility and establish the most efficient  correlation strategy, one needs a better 
approach than mean-field ones. The improved approximation should be able, unlike mean field, to disentangle 
charge, whose fluctuations are suppressed by a large $U$, from spin and orbital degrees of freedom which 
are not. For that purpose, we used a variational search within 
the class of Gutzwiller-type wave functions \cite{Gutzwiller_1,Gutzwiller_2}, much broader than Hartree-Fock which 
includes just Slater determinants and BCS wavefunctions.  In addition, we also adopted the so-called Gutzwiller 
approximation (GA) to evaluate the average values of any operator on the Gutzwiller wave function, an approximation 
that becomes exact in the limit of lattices with infinite coordination
\cite{bunemann1998,fabrizio2007,Fabrizio-Hvar}.

\subsection{Gutzwiller method}

The Gutzwiller variational wave function we shall consider is defined through 
\begin{equation}
\left|\Psi_{G}\right\rangle =\prod_{\bm{R}}\prod_{a=1,2}\, \mathcal{P}_{a\bm{R}}\, \left|\Psi_{0}\right\rangle ,
\end{equation}
where $ \mathcal{P}_{a \bm{R}}$ is a linear operator that depends on a set of variational parameters and acts on the LUMO+1 Hilbert space of molecule $a=1,2$ in the unit cell $\bR$, while $\left|\Psi_{0}\right\rangle$ is a variational Slater determinant or BCS wave function. $P_{a \bm{R}}$ therefore provides the new variational freedom with respect to Hartree-Fock. We impose the following pair of constraints  
\cite{bunemann1998,fabrizio2007,Fabrizio-Hvar}:
\begin{align}
\left\langle \Psi_{0}\right| \mathcal{P}_{a \bm{R}}^\dagger\, 
\mathcal{P}_{a \bm{R}}^\dagga \left|\Psi_{0}\right\rangle = & 1,\label{eq:norma}\\
\left\langle \Psi_{0}\right| \mathcal{P}_{a \bm{R}}^\dagger\, \mathcal{P}_{a \bm{R}}^\dagga \, 
n_{a\bm{R}\sigma}\left|\Psi_{0}\right\rangle = & \left\langle \Psi_{0}\right|n_{a\bm{R}\sigma}\left|\Psi_{0}\right\rangle,\label{eq:number}
\end{align}
where $n_{a\bR\sigma}$ is  the number operator of spin-$\sigma$ electrons on the LUMO+1 
of molecule $a$ at site $\bR$. Within the GA, and upon enforcing the above constraints, the following expressions are assumed, which 
are exact in infinite-coordination lattices, 
\ba
\left\langle \Psi_{G}\right|\hat{O}_{a \bm{R}}\left|\Psi_{G}\right\rangle &=& 
\left\langle \Psi_{0}\right|\,  \mathcal{P}_{a\bm{R}}^\dagger \, \hat{O}_{a \bm{R}}\,  
\mathcal{P}_{a \bm{R}}^\dagga \left|\Psi_{0}\right\rangle ,\\
\left\langle \Psi_{G}\right|\hat{O}_{a \bm{R}}\, \hat{O}_{b \bm{R}^{\prime}}\left|\Psi_{G}\right\rangle &=&\left\langle \Psi_{0}\right| \mathcal{P}_{a \bm{R}}^\dagger \, \hat{O}_{a\bm{R}}\,  
\mathcal{P}_{a \bm{R}}^\dagga \\ 
&& \qquad\qquad  \mathcal{P}_{b \bm{R}^{\prime}}^\dagger \, \hat{O}_{b \bm{R}^{\prime}}\, 
 \mathcal{P}_{b \bm{R}^{\prime}}^\dagga \left|\Psi_{0}\right\rangle ,
\ea
where $\hat{O}_{a \bm{R}}$ is any local operator. The right hand sides of both equations can be simply evaluated using 
Wick's theorem, which holds both for Slater determinants and BCS wave functions.  

The Hamiltonian we shall employ from now on is a further simplification of the original one in Sec. \ref{THE MODEL}.  
We already noticed that, among the two pairing channels, Eqs. \eqn{inter-molecule-pairing} and \eqn{pair-hopping}, 
only the first is able to circumvent a strong on-site repulsion. Therefore, we dismiss the pair hopping \eqn{pair-hopping} and approximate 
the phonon-mediated electron-electron interaction by  the inter-site pairing 
 \eqn{inter-molecule-pairing}, which we rewrite as
\bea
\mathcal{H}_\text{ep-eff} \rightarrow \mathcal{H}_\text{J} &=& 
J\,\sum_{\bR}\,\mathbf{S}_{1\bR}\cdot\mathbf{S}_{2\bR}\nonumber \\
&& + J\,\sum_{\bR}\,\mathbf{S}_{1\bR}\cdot\mathbf{S}_{2\bR+\mathbf{b}},\label{H-J}
\eea
where $\mathbf{S}_{a\bR}$ is the spin-operator of molecule $a=1,2$ at site $\bR$ and 
the second sum is restricted to nearest-neighbor molecules on different cells along the 
$\mathbf{b}$ direction. Here, $J$ has the same magnitude of $g_*$ in Eq. \eqn{el-el-phon-media}, 
and, for the sake of simplicity, we 
ignore
the retardation effects brought in by the functions 
$s_\bk$ in \eqn{el-el-phon-media}. This actually implies an underestimate of superconductivity, by not allowing for high-energy screening. 
Equation \eqn{H-J} also omits additional charge attraction between the molecules, which does not play any relevant role for large 
$U$\cite{Plekhanov-tJU}.  
The total simplified Hamiltonian then reads as
\begin{equation}
\mathcal{H}=\mathcal{H}_{0}+\mathcal{H}_{\mathrm{U}}+\mathcal{H}_{\mathrm{J}},\label{eq:tJ}
\end{equation}
and is 
a two-band version of the so-called $t$-$J$-$U$ model sometimes used in the context of 
high-$T_c$ superconductors\cite{Zhang-tJU,Plekhanov-tJU}, even though $J$ is provided here by 
electron-phonon coupling and not by projecting a purely electronic Hamiltonian  onto low-energy 
Zhang-Rice singlets of doped CuO$_2$ planes\cite{Zhang-Rice}.

We shall consider two possible variational wave-functions: a superconducting (SC) and an antiferromagnetic (AF) one. In the AF
state, molecule 1 is $\up$ spin-polarized and molecule 2 is 
$\down$ spin-polarized and the corresponding "uncorrelated" wave function $|\Psi_0\rangle$ 
is characterized by
\bea
\langle \Psi_0|\,n_{1\,\bR\up}-n_{1\,\bR\down}\,|\Psi_0\rangle &=& - \langle \Psi_0|\,n_{2\,\bR\up}-n_{2\,\bR\down}\,|\Psi_0\rangle \nonumber \\ && = 2m,\qquad \forall \bR
.\label{AF-m}
\eea
By contrast, the SC state has a spin-singlet intermolecular pairing as in 
Eq. \eqn{1-2-pairing}, which we assume real, and its uncorrelated wave function is thus characterized by
\be
\langle \Psi_0|\,c^\dagger_{1\bR\up}c^\dagger_{2\bRp\down}
+ c^\dagger_{2\bRp\up}c^\dagger_{1\bR\down}\,|\Psi_0\rangle = \Delta^{(0)}_{\bR\bRp}.
\label{SC-Delta}
\ee

 In both SC and AF cases, the linear operator $ \mathcal{P}_{a\bR}$ can be generally written as 
\bea
 \mathcal{P}_{a\bR} &=& \lambda_0\,|0_{a \bR}\rangle\langle 0_{a \bR}| + 
\lambda_2\,|2_{a \bR}\rangle\langle 2_{a \bR}| \nonumber\\
&& + \lambda_{a\,\up}\, |\up_{a \bR}\rangle\langle \up_{a \bR}| 
+ \lambda_{a\,\down}\, |\down_{a \bR}\rangle\langle \down_{a \bR}|,
\eea
in terms of projectors onto states with well-defined occupancies and spin of the LUMO+1 orbital of molecule $a$ at 
site $\bR$, 0 standing for empty, 2 for doubly occupied, and $\up(\down)$ for singly occupied with 
$\up(\down)$ spin. 
The variational parameters collectively designated as $\lambda$, which we can restrict to be real in this 
specific case, do not depend on the unit cell $\bR$ because we assume full translational symmetry. 
In addition, 
we shall not consider any charge disproportionation between the two molecules, then 
$\lambda_0$ and $\lambda_2$ are independent of $a$. On the contrary, in the AF case we must allow for 
$\lambda_{1\,\up}=\lambda_{2\,\down}\not = \lambda_{1\,\down}=\lambda_{2\,\up}$ consistently with the 
antiferromagnetic ordering, while the spin-singlet SC obviously implies $\lambda_{1\,\up}=\lambda_{2\,\down} = \lambda_{1\,\down}=\lambda_{2\,\up}$. 

We introduce the uncorrelated probability distribution 
$P^{(0)}_{a\alpha} = \langle \Psi_0|\,|\alpha_{a\bR}\rangle\langle \alpha_{a\bR}|\,|\Psi_0\rangle$, 
which is independent of $\bR$ and where $\alpha=0,2,\up,\down$, and the correlated one 
$P_{a\alpha} = \langle \Psi_G|\,|\alpha_{a\bR}\rangle\langle \alpha_{a\bR}|\,|\Psi_G\rangle$, which 
is readily found to be 
\ba
P_{a \,0} &=& \left|\lambda_{0}\right|^2\,P^{(0)}_{a \,0} = 
\frac{1}{4}\; \left|\lambda_{0}\right|^2\equiv P_0,\\
P_{a \,2} &=& \left|\lambda_{2}\right|^2\,P^{(0)}_{a \,2} = 
\frac{1}{4}\; \left|\lambda_{2}\right|^2\equiv P_2,\\
P_{1 \,\up} &=&  \left|\lambda_{1\,\up}\right|^2\,P^{(0)}_{1\,\up} 
=  \left|\lambda_{1\,\up}\right|^2\,\Big(\frac{1}{2}+m\Big)^2\\
&=& P_{2 \,\down} \equiv P_\up,\\
P_{1 \,\down} &=&  \left|\lambda_{1\,\down}\right|^2\,P^{(0)}_{1\,\down} 
=  \left|\lambda_{1\,\down}\right|^2\,\Big(\frac{1}{2}-m\Big)^2\\ 
&=& P_{2 \,\up} \equiv P_\down,
\ea
valid for both AF  ($m\not=0$) and SC ($m=0$).

Using these definitions, the constraints \eqn{eq:norma} and \eqn{eq:number}  
for our assumed density corresponding to one electron per site (three electrons per molecule, but
only one in the LUMO+1 orbital)  take the simple form
\bea
P_0 + P_\up + P_\down + P_2 &=& 1,\label{1}\\
P_0 &=& P_2,\label{2}\\
P_\up - P_\down &=& 2m,\label{3}
\eea
where in the SC wave function $P_\up=P_\down$ since $m=0$. 

The average energy of the variational wave function and within the GA is\cite{fabrizio2007} 
\bea
E &=& R^2\,\langle \Psi_0|\,\mathcal{H}_0\,|\Psi_0\rangle \nonumber\\
&& + \langle \Psi_0|\,\mathcal{H}_{*\,\text{J}}\,|\Psi_0\rangle + 2\,N\,P_0\,U,\label{E-1}
\eea
where $N$ is the number of unit cells,  
\be
R^2 = \fract{4 P_2}{1-4m^2}\;\Big(\sqrt{P_\up}+\sqrt{P_\down}\Big)^2,\label{R}
\ee
is a factor that renormalizes downwards the intersite hopping, (a factor whose square can be associated 
with the quasi-particle wave-function renormalization commonly denoted as $Z$), 
and $\mathcal{H}_{*\, \text{J}}$ has the same form of $\mathcal{H}_\text{J}$ provided the spin 
operators are modified according to 
$ \mathbf{S}_{a\bR} \rightarrow \mathcal{P}_{a\bR}\, \mathbf{S}_{a\bR} \,\mathcal{P}_{a\bR}$, 
which implies that (we omit for convenience the unit cell index $\bR$): 
\ba
\mathcal{P}_{a}\,S^z_{a}\,\mathcal{P}_{a}\, &=& 
\fract{\lambda_{a\, \up}^2 + \lambda_{a\, \down}^2}{2}\; S^z_a\\
&& + \fract{\lambda_{a\, \up}^2 - \lambda_{a\, \down}^2}{2}\;\Big(n_{a\up}+n_{a\down}
-2n_{a\up}n_{a\down}\Big),\\
\mathcal{P}_{a}\,S^+_{a}\,\mathcal{P}_{a}\, &=& \lambda_{a\,\up}\,\lambda_{a\,\down}\,S^+_{a}.
\ea
In the SC case one finds simply that $\mathcal{H}_{*\,\text{J}} = (4\,P_\up)^2\,\mathcal{H}_{\text{J}}=(4\,P_\down)^2\,\mathcal{H}_{\text{J}}$,  
showing that $J$ is renormalized to an effective $J_*= (4\,P_\up)^2\,J$, while in the AF case the expression becomes more involved.  

The variational energy in Eq. \eqn{E-1} depends on the $\lambda$ parameters, subject to the constraints of Eqs. \eqn{1}, \eqn{2}, and\eqn{3}.
It also depends on the 
uncorrelated wave function $|\Psi_0\rangle$, which is in turn constrained to the $\lambda$ only in the AF 
case through Eq. \eqn{AF-m}. Here, we find it more convenient to treat $m$ as an additional variational parameter, 
imposing Eq. \eqn{AF-m} via a Lagrange multiplier and \eqn{3} by simply 
setting $P_\up=2m+P_\down$ with $2P_\down=1-2P_0-2m\geq 0$, and minimizing first 
with respect to $|\Psi_0\rangle$ and $P_0$ and finally to $m$. We observe that optimization with respect to 
$|\Psi_0\rangle$ amounts to find either the antiferromagnetic Slater determinant, 
subject to the constraint \eqn{AF-m}, or the BCS wave function, both of which minimize the average value of 
$\mathcal{H}_*=\mathcal{H}_0+\mathcal{H}_{*\,\text{J}}$. This is practically the same task as solving a Hartree-Fock problem. 
The following step, i.e., minimization with respect to $P_0$ and, in the AF case, to $m$, is similarly
simple to accomplish, so that the full numerical optimization does not require a much greater effort than simple Hartree-Fock.

\begin{figure}
\includegraphics[scale=0.5,width=1.\linewidth]{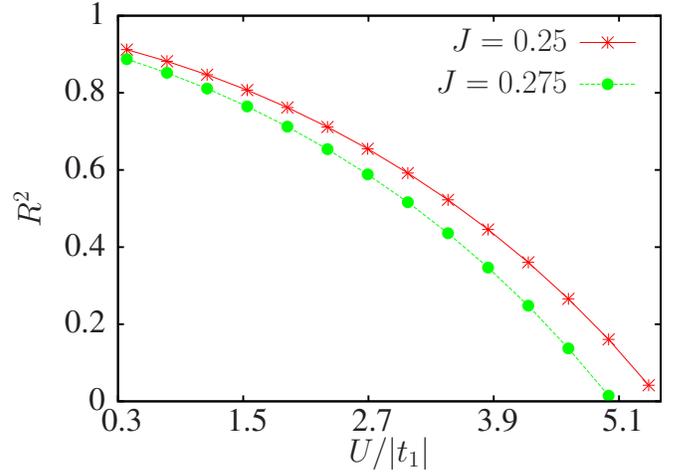}

\caption{\label{GA_R_SC}(Color online) Quasi-particle wave function renormalization factor $R^2$ of the superconducting state 
as function of $U$ in units of $|t_1|$. Note that $R^2$ goes continuously to 0, signaling a 
second-order Mott transition.}

\end{figure}

\begin{figure}
\includegraphics[scale=0.5,width=1.\linewidth]{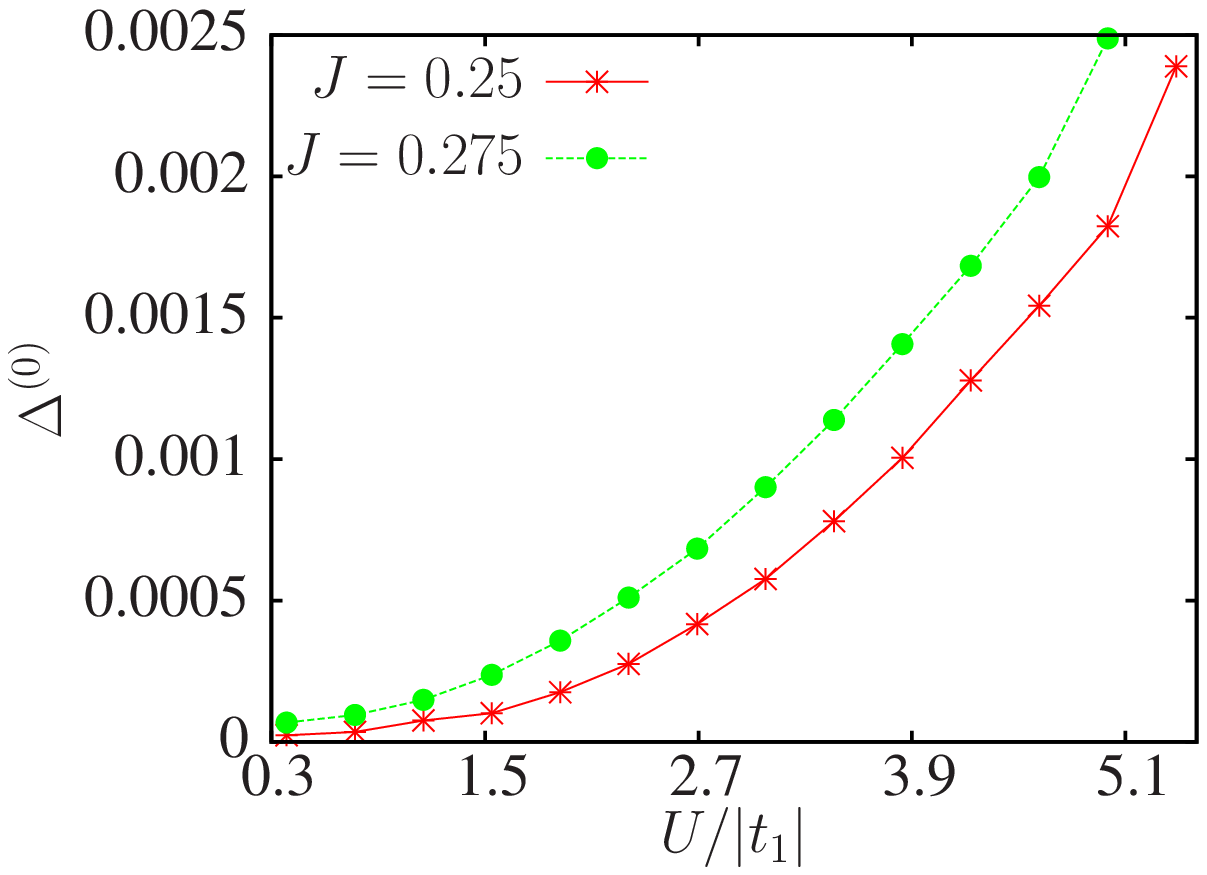}
\includegraphics[scale=0.5,width=1.\linewidth]{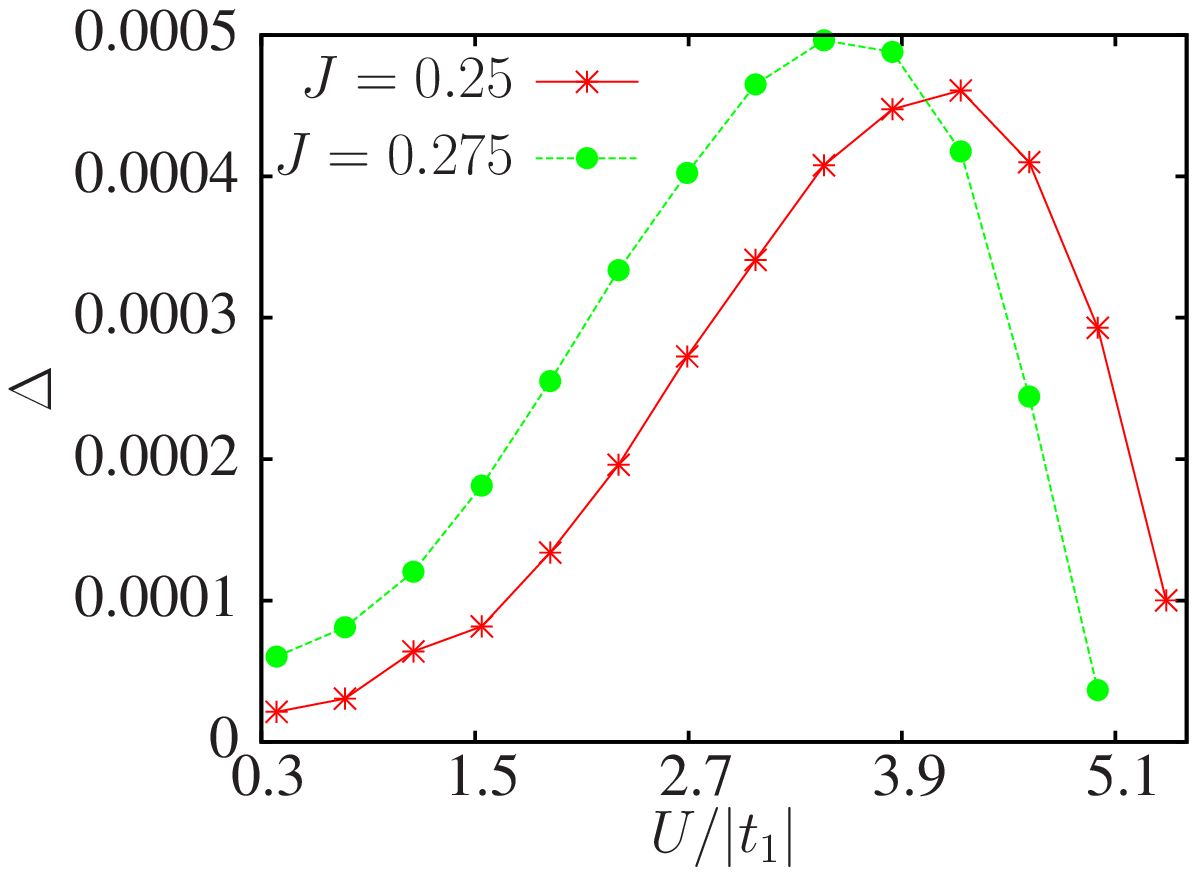}

\caption{\label{GA_SC_U}(Color online) Top panel:  uncorrelated SC parameter $\Delta^{(0)}$ as function of $U$. Lower panel:  true SC order parameter $\Delta$ as function of $U$. }

\end{figure}


\begin{figure}
\includegraphics[scale=0.5,width=1.\linewidth]{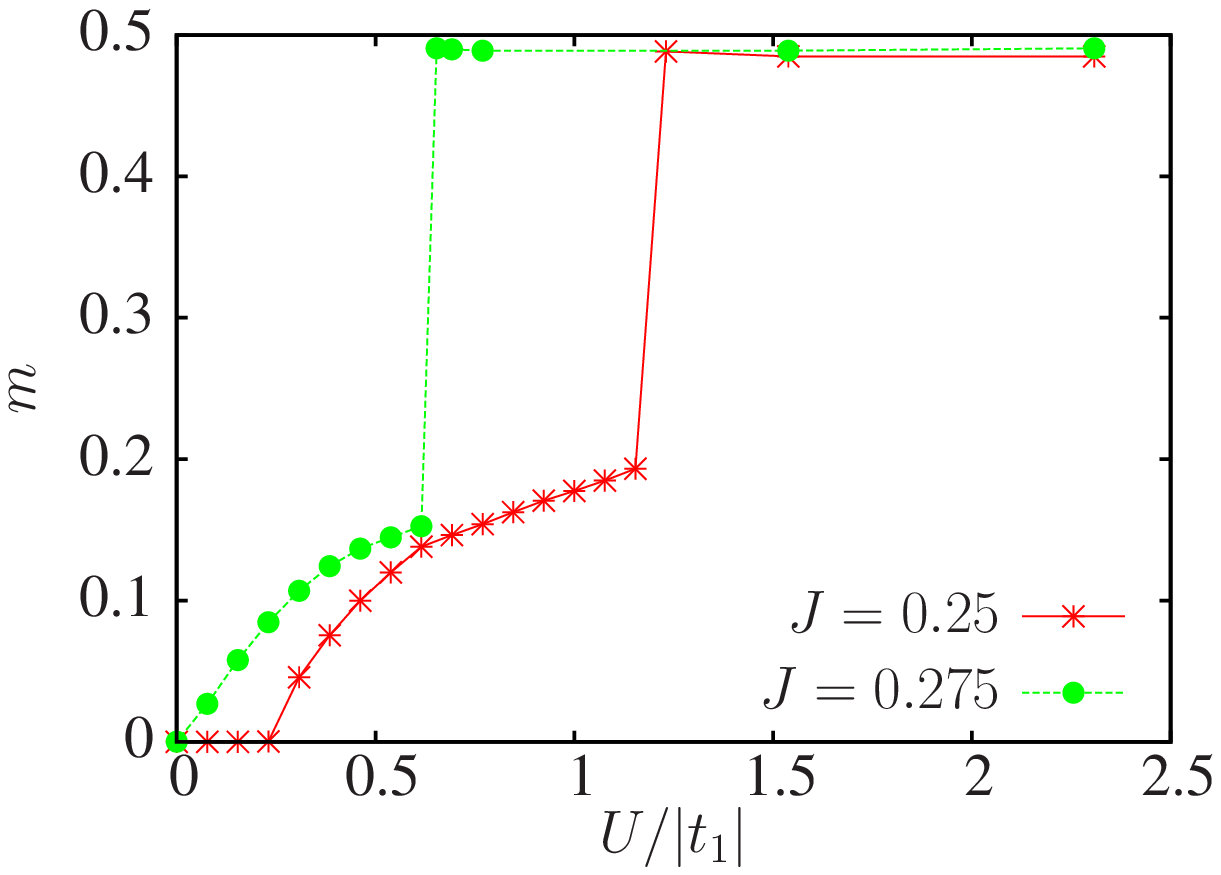}
\includegraphics[scale=0.5,width=1.\linewidth]{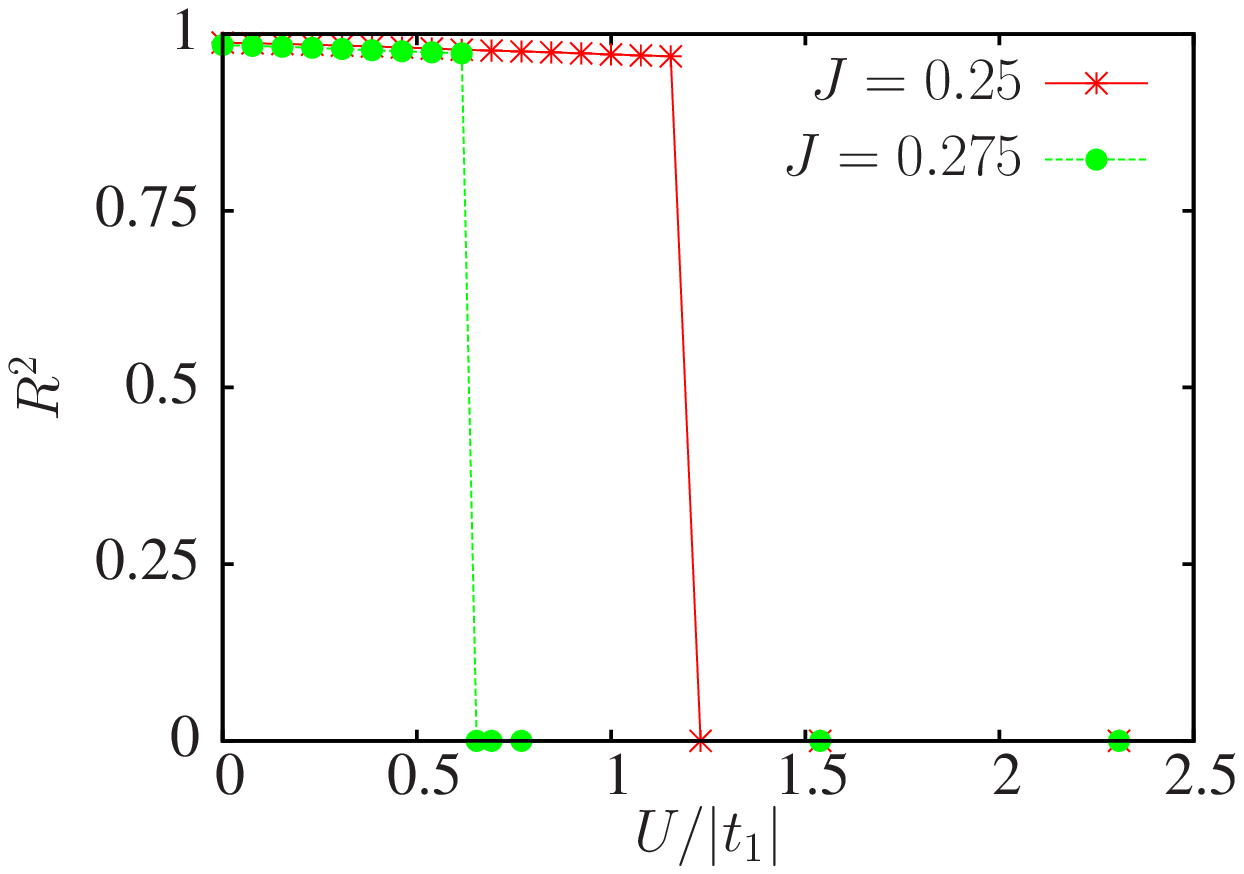}
\caption{\label{fig:AFM-dm}(Color online) Sublattice magnetization magnitude $m$ and renormalization factor $R^2$ of the antiferromagnetic
state with $U/|t_1|$. With increasing $U$, $m\rightarrow \frac{1}{2}$ and the system reaches the atomic limit. The metal-Mott insulator transition is of first order. }

\end{figure}

\begin{figure}
\includegraphics[scale=0.5,width=1.\linewidth]{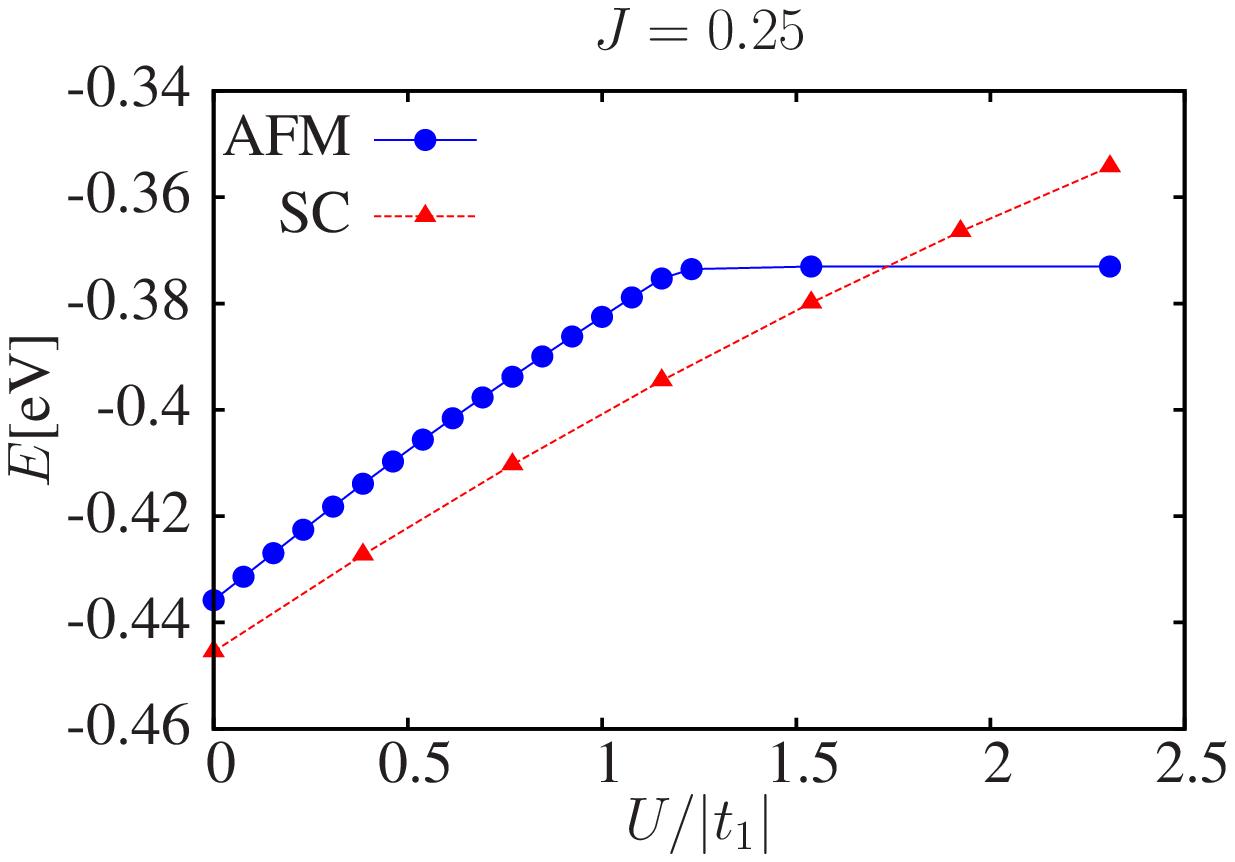}
\includegraphics[scale=0.5,width=1.\linewidth]{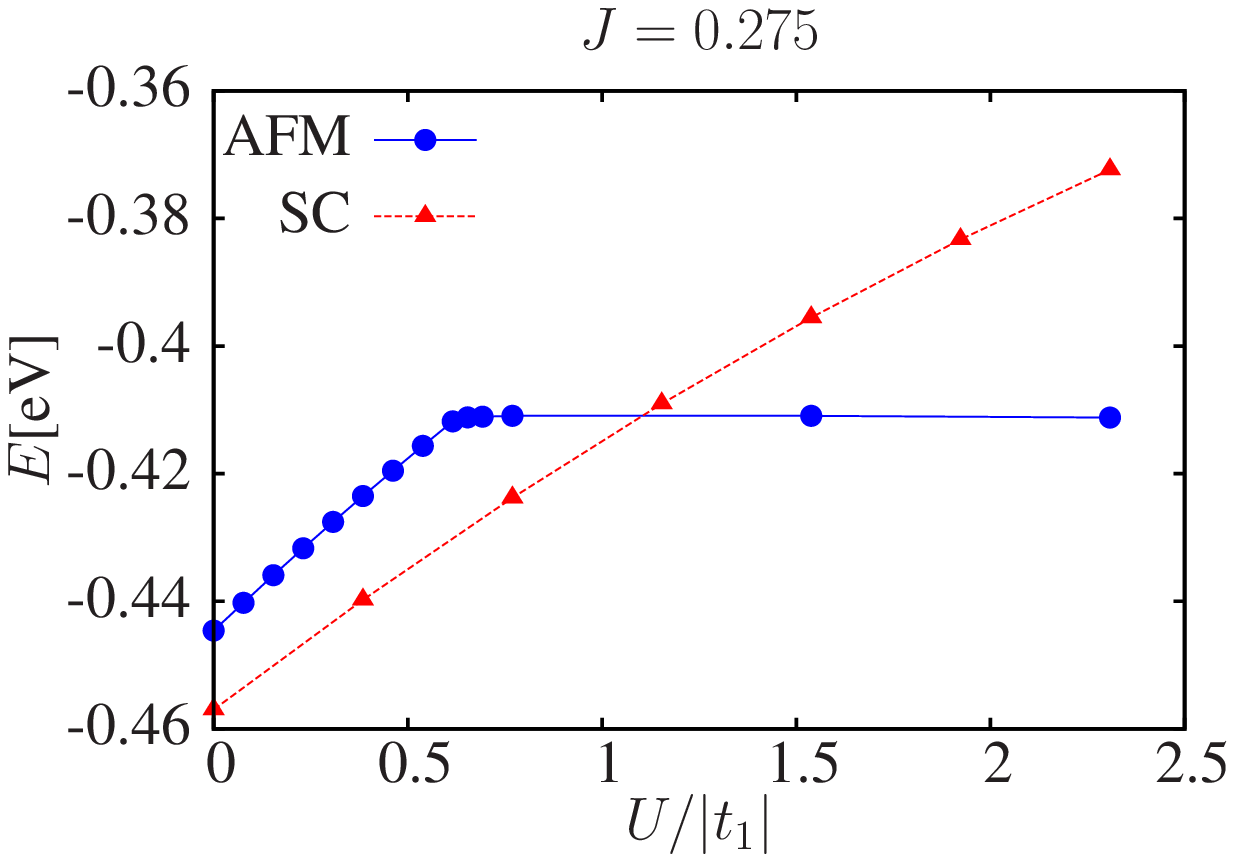}
\caption{\label{fig:energySC-AFM}(Color online) Competition between ground-state energies
of superconducting and antiferromagnetic states. Unlike mean field, where antiferromagnetism
always prevails, superconductivity survives with Gutzwiller correlations where there is a SC-AF transition  
at $U/|t_1|\simeq 1.8$ for $J=0.25$, and $U/|t_1|\simeq 1.1$ for $J=0.275$.}

\end{figure}

\subsection{Results and discussion}
We are now ready to present and discuss the results of the Gutzwiller variational approach for both SC and AF wave functions, starting from the former. 
We first note that the SC order parameter of the uncorrelated wave function, defined in Eq. \eqn{SC-Delta} no longer coincides 
with the actual order parameter evaluated on $|\Psi_G\rangle$. In fact, 
\bea
\Delta_{\bR\bRp}&\equiv& \langle \Psi_G|\,c^\dagger_{1\bR\up}c^\dagger_{2\bRp\down}
+ c^\dagger_{2\bRp\up}c^\dagger_{1\bR\down}\,|\Psi_G\rangle \nonumber \\
&& = R^2\, \Delta^{(0)}_{\bR\bRp},
\label{SC-Delta-true}
\eea
where the last expression derives from the GA. Since large $U$ suppresses double occupancy, that is 
$P_2=P_0\ll 1$, the renormalization factor $R\ll 1$. It is therefore possible to find an uncorrelated wave function 
$|\Psi_0\rangle$ with a large bare SC order parameter yet with just a tiny true order parameter after projection.  As we shall see, 
this is the key feature of the Gutzwiller wave function that allows the stabilization of superconductivity 
despite a strong repulsion, in fact even in the vicinity of a Mott transition, and has been repeatedly invoked in the 
context of $t$-$J$ models for cuprates\cite{Anderson-RVB,Randeria-t-J,sorella-t-J}. 

In Fig. \ref{GA_R_SC} we plot the wave-function renormalization factor $R^2$ as function of 
$U$. As expected, $R^2$ decreases monotonically with increasing $U$ and 
vanishes at a critical value that identifies the Mott transition. On the other hand, the inter-molecule SC order parameter 
$\Delta_0$ of the uncorrelated wave function increases with 
increasing 
$U$ (see top panel in 
Fig. \ref{GA_SC_U}). The joint result of these two variations is a non-monotonic variation of the true SC order parameter 
$\Delta$ [Eq. \eqn{SC-Delta-true}], which first increases with $U$, reaches a maximum, and then drops and vanishes at the Mott transition ( 
see lower panel in the same Fig. \ref{GA_SC_U}). This detailed behavior contrasts with the insensitivity of the Hartree-Fock mean-field 
solution of previous sections (see
Fig.\,\ref{MFA_gapU}). 

That at first sight surprising behavior  is actually explained by a physics not dissimilar to
that invoked  Refs. \onlinecite{capone2002} and \onlinecite{capone09} to explain the robustness of 
intra-molecular $s$-wave superconductivity in alkali-doped fullerenes in spite of the strong Coulomb repulsion. 
As $U$ increases, the effective quasi-particle bandwidth is renormalized down by the factor 
$R^2<1$. At the same time, the effective strength of the attraction $J_*=(4\,P_\up)^2\,J$ does not drop. 
In the present case of Fr{\"o}hlich intersite phonon pairing interaction $J_*$ actually increases, since 
the probability of single occupancy $P_\up=P_\down$ rises. The net result is that the system is pushed towards a strong 
coupling regime with an effective attraction of the same order of magnitude as the quasi-particle bandwidth. The reason 
why superconductivity is not damaged  but is fostered instead by an increasing repulsion,
is that, similarly to fulleride models, pairing in this model occurs in a channel orthogonal to charge\cite{capone2002,capone09}. 

We emphasize that these results are qualitatively not at all new, especially in the context $t$-$J$ models for cuprates. Indeed, 
exact calculations of Gutzwiller-projected wave functions, i.e., $P_2$ strictly zero but away from half-filling, using 
variational Monte Carlo already highlighted an increase of the uncorrelated order parameter upon approaching 
the undoped Mott insulator, as opposed to a reduction of the actual SC order parameter\cite{Randeria-t-J,sorella-t-J}.
However, these studies and calculations, which involved no phonons, dealt with 
$U\to \infty$, where metallic behavior is possible only at finite doping, whereas the half-filled system is trivially 
an antiferromagnetic Mott insulator. In our half-filled case, a correlated metallic and superconducting state is stable 
at small $U$, although it should eventually turn into an antiferromagnetic insulator above a critical $U_c$, which we must  
identify. For that purpose,  we implement the GA for the AF state, which is the natural competitor of SC. 

In Fig.\,\ref{fig:AFM-dm}, we plot the optimized staggered magnetization $m$ calculated as a function of $U$. 
We first observe that, because there are hopping processes that connect the same sublattice, the Fermi surface 
has no nesting hence a strong magnetization order parameter can only 
appears 
above a critical $U_*$ that diminishes by increasing $J$. 
We actually find {\emph {two}} different AF solutions, 
both insulating,  separated by a sharp transition. The first, characterized by a moderate staggered magnetization order parameter, 
is stable for small $U$. The second phase prevails above a threshold  value of $U$  where the magnetization jumps 
close to 
its maximum allowed value 0.5 and simultaneously the wavefunction renormalization $R^2$ drops to zero. 
We suspect that this transition between two insulating AF states is most likely an artifact of the GA approximation, 
which is known to describe strongly correlated insulators rather imperfectly. Indeed, as shown 
in Fig. \ref{fig:energySC-AFM}, the AF energy flattens out above the sharp transition, i.e. the insulating solution gets 
stuck into a state that does not change any more by further increasing $U$. 

At moderate $U$ values however, the Gutzwiller correlations are rather realistic.
In Fig. \ref{fig:energySC-AFM} we  compare the total energies of the correlated SC and AF 
optimized solutions, for 
moderate but 
increasing Coulomb repulsion $U$. Our main result is shown here. Correlations stabilize the SC state 
which now prevails over the AF solution in the whole 
region of small to moderate $U$ values. The prevalence of 
SC despite the local stability of an AF phase at small $U$ provides a strong measure of how effectively the Gutzwiller 
projection can suppress double occupancies out of the initial SC trial state.

\section{Conclusions}

In summary, a recently proposed Hubbard-Fr{\"o}hlich two-band model
is shown to possess an $s_\pm$ phonon-driven superconducting solution which,
in virtue of the cancellation due to the unlike sign of the two gaps,
can survive despite a sizable intrasite Hubbard repulsion. Upon inclusion of correlations, the order parameter 
may even actually benefit from an increasing $U$. The ground state remains superconducting for $U$
increasing from zero even if the antiferromagnetic solution exists as a locally stable energy minimum, until the two energies cross 
at a value of $U$ of the order half-bandwidth, and a first-order superconductor-to-antiferromagnetic insulator transition takes place.     
Previous models exhibiting opposite sign gaps were discussed
in particular by Agterberg \emph{et al.{} \cite{agterberg}} and by Mazin\cite{mazin2} in
the context of spin-fluctuations-driven superconductivity in iron pnictides,
where they are now under active consideration.  In our
case, a remarkable robustness of
$s_\pm$
superconductivity arises thanks
to  the near degeneracy of the two bands crossing the Fermi level in the normal metal, and by the
symmetry-breaking nature of the assumed phonon mode yielding a strong inter-site pairing. 

Because this model arose in the attempt to understand the as yet mysterious properties
of electron-doped PAHs, one could hope that it might be realized precisely
there. Should it become possible to create a Josephson junction of
a superconducting PAH compound and a regular BCS superconductor~\cite{agterberg},  the
model prediction would be amenable to direct test. 

\begin{acknowledgments}  Work supported  by  the  European  Union
FP7-NMP-2011-EU-Japan Project LEMSUPER, whose members are thanked for discussions.  
We also acknowledge MIUR Contract PRIN 2010LLKJBX\_004  and a CINECA  HPC award 2013.
\end{acknowledgments}

\section{Appendix\label{appendix}}

\begin{widetext}In this appendix, we derive $\mathcal{H}_{\mathrm{ep-eff}}$
by integrating 
out
the phonon 
degrees of
freedom. 

We start from the term $\mathcal{H}_{\mathrm{el-ph}}$. As defined
in Ref.~\onlinecite{naghavi}, we have $\gamma_{\bm{k}}\equiv\sum_{\mathbf{R}}g_{\left|\mathbf{R}-\bm{\delta}\right|}t_{\mathbf{R}}^{12}e^{-i\bm{k}\cdot\mathbf{R}}\left(R^{b}-\delta^{b}\right)$
where $\bm{\delta}=\left(\frac{1}{2},\,\frac{1}{2},\,0\right)$. The
following hopping is between molecules 1 and 2:
\begin{itemize}
\item $t_{1\mathbf{R}}^{12}=t_{1}$ for $\mathbf{R}=$(0, 0, 0 ) and (0, 1, 0),
hence $\mathbf{R}-\bm{\delta}=$($-\frac{1}{2}$, $-\frac{1}{2}$, $0$)
and ($-\frac{1}{2}$, $\frac{1}{2}$, 0), respectively;
\item $t_{2\mathbf{R}}^{12}=t_{2}$ for $\mathbf{R}=$(1, 0, 0) and (1, 1, 0), hence
$\mathbf{R}-\bm{\delta}=$ ($\frac{1}{2}$, $-\frac{1}{2}$, $0$) and
($\frac{1}{2}$, $\frac{1}{2}$, $0$), respectively;
\item $t_{3\mathbf{R}}^{12}=t_{3}$ for $\mathbf{R}=$(0, 0, -1) and (0, 1, -1),
hence $\mathbf{R}-\bm{\delta}=$(-$\frac{1}{2}$, $-\frac{1}{2}$, $-1$)
and ($-\frac{1}{2}$, $\frac{1}{2}$, $-1$), respectively.
\end{itemize}
So we have 
\begin{equation}
\gamma_{\bm{k}}=g\left(-1+e^{-ik_{y}}\right)\left(t_{1}+t_{2}e^{-ik_{x}}+t_{3}e^{ik_{z}}\right)=-ig\tan\frac{k_{y}}{2}t_{\bm{k}}^{12}=ig\tan\left(\frac{k_{y}}{2}\right)e^{-i\theta_{\bm{k}}}\tau_{\bm{k}}.
\end{equation}
Now we can write the el-ph coupling as
\begin{equation}
\mathcal{H}_{\mathrm{el-ph}}=-ig\sum_{\bm{k}\bm{k}^{\prime},\sigma}\left(x_{\bm{k}-\bm{k}^{\prime}}\Gamma_{\bm{k}^{\prime},\bm{k}}c_{1\bm{k}\sigma}^{\dagger}c_{2\bm{k}^{\prime}\sigma}-x_{\bm{k}-\bm{k}^{\prime}}\Gamma_{\bm{k},\bm{k}^{\prime}}^{\ast}c_{2\bm{k}\sigma}^{\dagger}c_{1\bm{k}^{\prime}\sigma}\right),\label{eq:ep}
\end{equation}
where $\Gamma_{\bm{k},\bm{k}^{\prime}}\equiv\tan\left(\frac{k_{y}}{2}\right)t_{\bm{k}}^{12}+e^{-i\phi_{\bm{k}-\bm{k}^{\prime}}}\tan\left(\frac{k_{y}^{\prime}}{2}\right)t_{\bm{k}^{\prime}}^{12}$
and we used $e^{i\phi_{\bm{q}}}\equiv\frac{1+e^{iq_{y}}}{\left|1+e^{iq_{y}}\right|}$. 

Combining Eq. (\ref{eq:ep}) and $\mathcal{H}_{\mathrm{ph}}$, and
integrating out the phonons, we find an effective contribution to
the action

\begin{align}
\delta\mathcal{S}= & \sum_{\bm{k}\bm{k}\sigma}D\left(\bm{k}-\bm{k}^{\prime},\epsilon-\epsilon^{\prime}\right)\left[\left|\Gamma_{\bm{k}^{\prime},\bm{k}}\right|^{2}\left(c_{1\bm{k}\sigma}^{\dagger}\left(\epsilon\right)c_{1-\bm{k}-\sigma}^{\dagger}\left(-\epsilon\right)c_{2-\bm{k}^{\prime}-\sigma}\left(-\epsilon^{\prime}\right)c_{2\bm{k}^{\prime}\sigma}\left(\epsilon^{\prime}\right)+\left(1\leftrightarrow2\right)\right)\right.\nonumber \\
 & -\Gamma_{\bm{k}^{\prime},\bm{k}}\Gamma_{\bm{k},\bm{k}^{\prime}}c_{1\bm{k}\sigma}^{\dagger}\left(\epsilon\right)c_{2-\bm{k}-\sigma}^{\dagger}\left(-\epsilon\right)c_{1-\bm{k}^{\prime}-\sigma}\left(-\epsilon^{\prime}\right)c_{2\bm{k}^{\prime}\sigma}\left(\epsilon^{\prime}\right)\nonumber \\
 & \left.-\Gamma_{\bm{k}^{\prime},\bm{k}}^{\ast}\Gamma_{\bm{k},\bm{k}^{\prime}}^{\ast}c_{2\bm{k}\sigma}^{\dagger}\left(\epsilon\right)c_{1-\bm{k}-\sigma}^{\dagger}\left(-\epsilon\right)c_{2-\bm{k}^{\prime}-\sigma}\left(-\epsilon^{\prime}\right)c_{1\bm{k}^{\prime}\sigma}\left(\epsilon^{\prime}\right)\right],
\end{align}
where $D\left(\bm{k},\epsilon\right)=-\frac{g^{2}\omega_{\bm{k}y}}{\epsilon^{2}+\omega_{\bm{k}y}^{2}}$.
Using the transformations in Eqs.~\eqref{c-1} and~\eqref{c-2}, and concentrating on the intra-band
pairing, we have 
\begin{align}
\delta\mathcal{S}= & \frac{1}{2}\sum_{\bm{k}\bm{k}^{\prime}\sigma}D\left(\bm{k}-\bm{k}^{\prime},\epsilon-\epsilon^{\prime}\right)\left[\left|\Gamma_{\bm{k}^{\prime},\bm{k}}\right|^{2}\left(\Delta_{g\bm{k}\sigma}^{\dagger}\left(\epsilon\right)+\Delta_{u\bm{k}\sigma}^{\dagger}\left(\epsilon\right)\right)\left(\Delta_{g\bm{k}^{\prime}\sigma}\left(\epsilon^{\prime}\right)+\Delta_{u\bm{k}^{\prime}\sigma}\left(\epsilon^{\prime}\right)\right)\right.\nonumber \\
 & \left.-\Re\left[\Gamma_{\bm{k}^{\prime},\bm{k}}\Gamma_{\bm{k},\bm{k}^{\prime}}e^{i\left(\theta_{\bm{k}}+\theta_{\bm{k}^{\prime}}\right)}\right]\left(\Delta_{g\bm{k}\sigma}^{\dagger}\left(\epsilon\right)-\Delta_{u\bm{k}\sigma}^{\dagger}\left(\epsilon\right)\right)\left(\Delta_{g\bm{k}^{\prime}\sigma}\left(\epsilon^{\prime}\right)-\Delta_{u\bm{k}^{\prime}\sigma}\left(\epsilon^{\prime}\right)\right)\right],
\end{align}
where $\Delta_{g\left(u\right)\bm{k}\sigma}^{\dagger}\left(\epsilon\right)=c_{g\left(u\right)\bm{k}\sigma}^{\dagger}\left(\epsilon\right)c_{g\left(u\right)-\bm{k}-\sigma}^{\dagger}\left(-\epsilon\right)$.
Further simplification leads to

\begin{align}
\delta\mathcal{S} & =\frac{1}{2}\sum_{\bm{k}\bm{k}^{\prime}\sigma}D\left(\bm{k}-\bm{k}^{\prime},\epsilon-\epsilon^{\prime}\right)\left[A_{\bm{k},\bm{k}^{\prime}}\left(\Delta_{g\bm{k}\sigma}^{\dagger}\left(\epsilon\right)\Delta_{g\bm{k}^{\prime}\sigma}\left(\epsilon^{\prime}\right)+\Delta_{u\bm{k}\sigma}^{\dagger}\left(\epsilon\right)\Delta_{u\bm{k}^{\prime}\sigma}\left(\epsilon^{\prime}\right)\right)\right.\nonumber \\
 & +B_{\bm{k},\bm{k}^{\prime}}\left(\Delta_{g\bm{k}\sigma}^{\dagger}\left(\epsilon\right)\Delta_{u\bm{k}^{\prime}\sigma}\left(\epsilon^{\prime}\right)+\Delta_{u\bm{k}\sigma}^{\dagger}\left(\epsilon\right)\Delta_{g\bm{k}^{\prime}\sigma}\left(\epsilon^{\prime}\right)\right),\label{eq:deltaS}
\end{align}
where 

\[A_{\bm{k},\bm{k}^{\prime}}=\left(1-\cos\left(\theta_{\bm{k}}-\theta_{\bm{k}^{\prime}}-\phi_{\bm{k}-\bm{k}^{\prime}}\right)\right)\left(\tan\left(\frac{k_{y}}{2}\right)t_{\bm{k}}^{12}-\tan\left(\frac{k_{y}^{\prime}}{2}\right)t_{\bm{k}^{\prime}}^{12}\right),\] and 

\[B_{\bm{k},\bm{k}^{\prime}}=\left(1+\cos\left(\theta_{\bm{k}}-\theta_{\bm{k}^{\prime}}-\phi_{\bm{k}-\bm{k}^{\prime}}\right)\right)\left(\tan\left(\frac{k_{y}}{2}\right)t_{\bm{k}}^{12}+\tan\left(\frac{k_{y}^{\prime}}{2}\right)t_{\bm{k}^{\prime}}^{12}\right).\]

The Fermi surface is close to $k_{y}=\pm\pi$. We can denote the right-moving
fermions (R) as those with $k_{y}=\pi-\kappa$ and left moving fermions
(L) as those with $k_{y}=-\pi+\kappa$, with $\pi\gg\kappa>0$. For
$\theta_{\bm{k}}=\tan^{-1}\left(-\tan\frac{k_{y}}{2}\right)+\varphi_{\bm{k}}$
where $\varphi_{\bm{k}}=\arctan\frac{t_{2}\sin k_{x}-t_{3}\sin k_{z}}{-t_{1}-t_{2}\cos k_{x}-t_{3}\cos k_{z}}$,
we have 
\begin{align}
\theta_{R\bm{k}}\simeq & -\frac{\pi}{2}+\varphi_{\bm{k}},\label{eq:R}\\
\theta_{L\bm{k}}\simeq & +\frac{\pi}{2}+\varphi_{\bm{k}},\label{eq:L}
\end{align}
with the phonon phase $\phi_{\bm{k}-\bm{k}^{\prime}}\simeq0$. We
can see that the first term in Eq. (\ref{eq:deltaS}) is nonzero only when
$\theta_{\bm{k}}=\theta_{R\bm{k}}$ and $\theta_{\bm{k}^{\prime}}=\theta_{L\bm{k}^{\prime}}$,
or $\theta_{\bm{k}}=\theta_{L\bm{k}}$ and $\theta_{\bm{k}^{\prime}}=\theta_{R\bm{k}^{\prime}}$.
We have $A_{\bm{k},\bm{k}^{\prime}}\simeq\left(1+\cos\left(\varphi_{\bm{k}}-\varphi_{\bm{k}^{\prime}}\right)\right)\left(\zeta_{\bm{k}}+\zeta_{\bm{k}^{\prime}}\right)^{2}$
where $\zeta_{\bm{k}}=2\left|\sin\frac{k_{y}}{2}\left(t_{1}+t_{2}e^{-ik_{x}}+t_{3}e^{ik_{z}}\right)\right|$.
For the first term in Eq. (\ref{eq:deltaS}), we observe that 
\begin{align}
c_{R\uparrow}^{\dagger}c_{L\uparrow}^{\dagger}c_{R\downarrow}c_{L\downarrow}+c_{L\uparrow}^{\dagger}c_{R\uparrow}^{\dagger}c_{L\downarrow}c_{R\downarrow} & =\frac{1}{2}\left(c_{R\uparrow}^{\dagger}c_{L\downarrow}^{\dagger}+c_{L\uparrow}^{\dagger}c_{R\downarrow}^{\dagger}\right)\left(c_{L\downarrow}c_{R\uparrow}+c_{R\downarrow}c_{L\uparrow}\right)\nonumber \\
 & -\frac{1}{2}\left(c_{R\uparrow}^{\dagger}c_{L\downarrow}^{\dagger}-c_{L\uparrow}^{\dagger}c_{R\downarrow}^{\dagger}\right)\left(c_{L\downarrow}c_{R\uparrow}-c_{R\downarrow}c_{L\uparrow}\right).\label{eq:pair}
\end{align}
In Eq. (\ref{eq:pair}) we can see the singlet and triplet pairings.
Since the phonon mediated coupling $D\left(\bm{k}-\bm{k}^{\prime},\epsilon-\epsilon^{\prime}\right)$
is attractive at the low frequency, it only favors the singlet pairing.
Thus the first term in Eq. (\ref{eq:deltaS}) is 
\[
\left(1+\cos\left(\varphi_{\bm{k}}-\varphi_{\bm{k}^{\prime}}\right)\right)\left(\zeta_{\bm{k}}+\zeta_{\bm{k}^{\prime}}\right)^{2}\left(\Delta_{g\bm{k}}^{\dagger}\left(\epsilon\right)\Delta_{g\bm{k}^{\prime}}\left(\epsilon^{\prime}\right)+\Delta_{u\bm{k}}^{\dagger}\left(\epsilon\right)\Delta_{u\bm{k}^{\prime}}\left(\epsilon^{\prime}\right)\right).
\]
Finally neglecting the momentum dependence in the coefficient, we recover
Eq.~\eqref{el-el-phon-media} in the main text. 

The second term in Eq. (\ref{eq:deltaS}) is nonzero only
when $\theta_{\bm{k}}=\theta_{R\bm{k}}$ and $\theta_{\bm{k}^{\prime}}=\theta_{R\bm{k}^{\prime}}$,
or $\theta_{\bm{k}}=\theta_{L\bm{k}}$ and $\theta_{\bm{k}^{\prime}}=\theta_{L\bm{k}^{\prime}}$.
We observe that in Eq. (\ref{eq:deltaS}) these
two terms cannot exist at the same time. When calculating the gap
parameters adopting the second term in Eq. (\ref{eq:deltaS}), the gap
symmetry does not change. 

\end{widetext}


%

\end{document}